\documentclass[aps,superscriptaddress,showpacs,showkeys,a4paper, 10pt,nofootinbib,notitlepage]{revtex4}

\usepackage[utf8]{inputenc}
\usepackage{amsmath}
\usepackage{amssymb}
\usepackage[english]{babel}
\usepackage{graphics}
\usepackage{graphicx}
\usepackage{epsfig}
\def\vec#1{\mbox{\bf #1}}              
\def\vecs#1{\mbox{\scriptsize \bf #1}} 
\renewcommand{\vec}{\boldsymbol}

\allowdisplaybreaks[2]

\begin{document}

\title{Spontaneous emission in a quantum system driven by the resonant field beyond the rotating wave approximation}

\author{I.\ D.\ Feranchuk}
\email[Corresponding author: ]{ilya_feranchuk@tdt.edu.vn}
\affiliation{Atomic Molecular and Optical Physics Research Group, Ton Duc Thang University, 19 Nguyen Huu Tho Str., Tan Phong Ward, District 7, Ho Chi Minh City, Vietnam}
\affiliation{Faculty of Applied Sciences, Ton Duc Thang University, 19 Nguyen Huu Tho Str., Tan Phong Ward, District 7, Ho Chi Minh City, Vietnam}
\affiliation{Belarusian State University, 4 Nezavisimosty Ave., 220030, Minsk, Belarus}
\author{A.\ U.\ Leonau}
\email[Corresponding author: ]{leonov.bsu@gmail.com}
\affiliation{Belarusian State University, 4 Nezavisimosty Ave., 220030, Minsk,   Belarus}
\author{M.\ M.\ Eskandari}
\email[Corresponding author: ]{cosmic.mahdi@gmail.com}
\affiliation{Institute of Physics of National Academy of Sciences, 68 Nezavisimosty Ave., 220068, Minsk,   Belarus}

\begin{abstract}
Quasi-stationary states of the quantum system in the driving resonant field are considered without rotating wave approximation. Conditions under which the spontaneous emission could be suppressed in this system are investigated in the special case when the frequency of the driving field is essentially less than the frequency of spontaneous emission. It is shown that the characteristic parameters of the effect are substantially changed  in comparison with the results corresponding to RWA. The real physical system for which the considered effects could be observed is described.
\end{abstract}

\maketitle

\section{Introduction.}

It is supposed in most of the papers considering the relaxation processes in quantum optics that the spontaneous (natural) widths of the atomic levels are their intrinsic characteristics and are fixed \cite{Agarval, Scully}. However, it was shown in several papers  that these values can be controlled by means of a low-frequency driving field.  Some of the first theoretical predictions of this phenomenon were published in \cite{2} and \cite{3}.  In \cite{4}  its detailed description  was presented for a two-level atom. Recently,  the first experimental demonstration of the fluorescence spectral line elimination in a quantum system induced by the interfering transition amplitudes of the relaxing states    was reported \cite{1}.

In the majority of papers  the "dressed" states of a radiating two-level system exposed  to  a laser field are described within the framework of the Rotating Wave Approximation (RWA). This approximation proves to be valid in the case of rather weak atom-field interaction. In several  papers the spontaneous emission was considered beyond RWA, however "anti-rotating" terms  in the Hamiltonian of the system were considered as small values and were taken into account in the framework of the perturbation theory \cite{nonRWA_1996, nonRWA_2011}.

The goal of the present paper is to show that in the case of the strong field these terms lead to essential changes of the dipole matrix element of the two level atom and the parameters of its spontaneous emission. Besides, we describe the method for calculating  these parameters and analysing the spontaneous emission without the perturbation theory. Our approach is based on the Uniformly Available Approximation (UAA) for the states of a two-level atomic system in a single mode resonance quantum field that was found in the analytical form in \cite{5} and \cite{6}.    UAA  appeared to be valid  in the whole range of the varying amplitude of the resonance field  in  such systems \cite{7,8}.
It was shown  in the paper \cite{5}  and described later in \cite{6,7,8,8b}  that RWA  applicability is defined not only by the atom-field coupling constant $f$ but also by the quantum number $n$ of the field according to inequality $2 f \sqrt{n} \ll 1$. In the case of zero detuning value this parameter coincides with Rabi frequency $\Omega_R$. It was shown in \cite{8b} that UAA can also be  used in the case of nonzero detuning value and RWA  applicability is defined by the same parameter.

Thus, according  to \cite{5,6,8, 8b}, the RWA is applicable if the following condition is fulfilled:

\begin{eqnarray}
\label{1}
    \Omega_R \ll \Omega,
\end{eqnarray}

\noindent where $\Omega$ is the frequency of the resonant field and $\Omega_R$ is the Rabi frequency of the two-level system. It is important to stress that the case $\Omega_R \sim \Omega$ can exist for real physical systems, and the transition from weak to strongly driven regime has been discovered experimentally \cite{8a}.

On the other hand, the dynamical control  of  the processes of the spontaneous emission in a two-level system becomes   effective when the lifetime of the spontaneous decay is  bigger than the period of Rabi oscillations. This condition is defined by inequality  \cite{Agarval,Scully}: 

\begin{eqnarray}
\label{2}
    \Omega_R >   \Gamma,
\end{eqnarray}

\noindent where $\Gamma$ is the spontaneous emission linewidth of the two-level system (here and below we adopt the system of units with $\hbar = c = 1$).

Thus, for the characteristic parameters of the system ($\Gamma , \Omega_R , \Omega$) there exists a range

\begin{eqnarray}
\label{2a}
\Gamma < \Omega_R \leq \Omega,
\end{eqnarray}

\noindent where the   counter-rotating terms in the two-level system Hamiltonian should be included when calculating  characteristics of the spontaneous emission and fluorescence spectra.   This allows one to analyze the fluorescence spectra  correctly  and estimate  the  field parameters for  the efficient  control over the evolution of the system.
Analysis  of the role of such "corrections" in the formation and control of the spontaneous emission spectra  is the main goal of the present paper.

The paper is organized in the following way. In Sec.\ref{sec.2}  the simple atomic model is considered as an example in order to illustrate qualitatively how one can use UAA instead of RWA for the description of the atomic "dressed" states. In Sec.\ref{sec.3}  the non-RWA effects in the control of the spontaneous emission processes are analyzed for the same model. In Sec.\ref{sec.4} the real physical system is proposed which could allow one to observe these effects.

\section{Description of the sample model  }
\label{sec.2}

\begin{figure}[tb]
\includegraphics[scale=0.5]{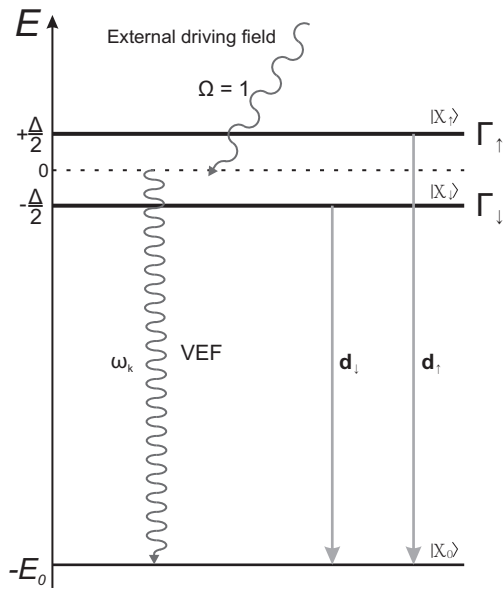}
\caption{The basic scheme of the V-configuration three-level model.}
\label{Fig.1}
\end{figure}

Analysis of evolution of the quantum systems in the resonant quantum field beyond the RWA is not widely spread. Therefore, let us consider first of all rather simple illustrative model (V-configuration three-level system, see Fig.\ref{Fig.1}) in order to make it clear how one can use UAA in this case.

As it is known \cite{10},  the  interaction between the atom and the vacuum electromagnetic field (VEF) leads to the  formation of  a radiation width of quantum levels. This width is defined by the spontaneous emission and can be calculated  via  the quasi-stationary solution of the Schr\"odinger equation:

\begin{eqnarray}
\label{3}
\hat H | \psi \rangle = E | \psi \rangle; \nonumber \\
\hat H = \hat H_a + \hat H_r + \hat H_e + \hat V_r + \hat V_e.
\end{eqnarray}

For the three-level configuration (Fig.\ref{Fig.1}) the total Hamiltonian of the system in the nonrelativistic case  is composed of the following operators:

\begin{eqnarray}
\label{4}
\hat H_a = - E_0 | \chi_0 \rangle
\langle \chi_0 | + \frac{1}{2} \Delta \sigma_3;\nonumber \\
\hat H_e =  a^+ a; \qquad \hat V_e = f \sigma_1 (a + a^+);
\end{eqnarray}

\begin{eqnarray}
\label{5}
\hat H_r = \sum_{\vecs{k},s} \omega_{\vecs{k}}
b^+_{\vecs{k},s} b_{\vecs{k},s};  \
\hat V_r =
- \frac{e}{m_e} \sum_{\vecs{k},s}
(\frac{2\pi}{V \omega_{\vecs{k}}})^{1/2} e^{i\vecs{k}\vecs{r}}
(\vec{e}_{\vecs{k},s} \hat{\vec{p}})
(b^+_{-\vecs{k},s} + b_{\vecs{k},s}).
\end{eqnarray}

\noindent  We scale the energy system by the frequency of the external driving field so that $\Omega = 1$ could be used without the loss of generality.

Here  $(-E_0), E_0 > 0$   is the energy of the ground state (in   units  of $\Omega $ ) with its eigenstate  $|\chi_0 \rangle$. The energy splitting of the excited double-states $|\chi_{\uparrow,\downarrow}\rangle$   with the energies $\pm \frac{1}{2}\Delta$  is presented in the space of Pauli matrices $\sigma_i$; the coupling constant $f$ defines the resonant interaction between the atom and the external driving field, it is proportional to the dipole  momentum  of  the transition between the excited double-states; $a^{+}, a$  are the creation and annihilation operators  of the quanta of the external single-mode field;  $b^+_{\vecs{k},s}, b_{\vecs{k},s}$  are the creation and annihilation operators  of  the  VEF photon  with the wave vector $\vec{k}$ , frequency  $\omega_{\vecs{k}}$ and polarization $\vec{e}_{\vecs{k},s}$; $e, m_e$ and $ \hat{\vec{p}}$ are the charge, mass and the momentum operator of the atomic electron correspongingly, $V$ is the normalization volume. In this  system of units the Rabi frequency  corresponding to $n$-quantum state of the resonance field is defined as

\begin{eqnarray}
\label{5a}
\Omega_R = 2 f \sqrt{n}.
\end{eqnarray}

In order to take into account the interaction between the atom and the driving resonant filed,  let us define  a new basis set   for the excited states of the atom  that are  usually called the "dressed states" \cite{Scully}.
In general case the "dressed states" are  represented by the exact solutions of the equation:

\begin{eqnarray}
\label{6}
\{\frac{1}{2} \Delta \sigma_3 + \Omega a^+ a + f \sigma_1 (a + a^+)
- E^{(0)}_{np}\} | \Psi_{np} \rangle = 0.
\end{eqnarray}

Thus, the two excited states of the "bare" atom  are transformed to  the  spectra of states with the quasi-energies    $E_{np}$  depending  on the quantum number $n$  that corresponds to  the number of photons in the driving field and the quantum number $p = \pm 1$  that corresponds to the eigenvalue of the   parity operator commuting with the Hamiltonian of the two-level system  in the quantum field \cite{5}:

\begin{eqnarray}
\label{7}
\hat P | \Psi_{np} \rangle \equiv \sigma_3 e^{i\pi a^+ a}
| \Psi_{np} \rangle = p| \Psi_{np} \rangle.
\end{eqnarray}

In most papers, the solution of (\ref{6}) is obtained using RWA   which leads to a simple analytical expression for the spectra of quasi-energies \cite{Scully}. However, according to \cite{5}   RWA is applicable only for sufficiently small values $\Omega_R \ll 1$. Taking into account the range (\ref{2a}),  one can expect that the correction  in  the quasi-energies stipulated by the off-resonant terms in (\ref{6}) might be comparable with the radiation linewidth of the excited states. To  go beyond the RWA framework, an exact solution of Eqns.  (\ref{6}) and (\ref{7})  obtained  in \cite{5} can be used:

\begin{eqnarray}
\label{8}
| \Psi_{np} \rangle = \sum_{k=0}^{\infty}\sum_{q=\pm}
C^{np}_{kq} |k, f \rangle |\chi_q \rangle,
\end{eqnarray}

\noindent where  the expansion coefficients   $C^{np}_{kq}$  for  arbitrary quantum numbers and the coupling constant $f$ are calculated numerically by  means of  fast converging recurrence procedure which is based on the Operator Method (OM) \cite{12}.  According to OM, a set of the basic states is defined as follows:

\begin{eqnarray}
\label{9}
|n, f \rangle = \frac{(a^+ + f )^n}{\sqrt{n!}}
\sum_{k=0}^{\infty}\frac{f^k}{{k!}} (a^+)^k |0_a\rangle e^{-f^2/2};
\  a |0_a\rangle = 0; \
|\chi_{\pm} \rangle = \frac{1}{\sqrt{2}}
(|\chi_{\uparrow}\rangle \pm |\chi_{\downarrow} \rangle).
\end{eqnarray}

However, a more convenient way for practical calculations is defined by the analytical UAA representation for solutions of the Eqns. (\ref{6}) and (\ref{7}), which was  obtained  for the first time in \cite{5}  (see also \cite{7}). It is defined by the following expressions  for the  quasi-energies:

\begin{eqnarray}
\label{10}
E^{(0)}_{np} \simeq (n + \frac{1}{2}q)  -  f^2
+ \frac{1}{4} p \Delta (S_{nn} + S_{n+q, n+q}) - \frac{1}{2}qM;\nonumber\\
q = p (-1)^n; \quad M = \{ [1 - \frac{1}{2}\Delta (-1)^n
(S_{nn} - S_{n+q, n+q})]^2 + \Delta^2 S^2_{n, n+q}\}^{1/2};\nonumber\\
S_{km} = (-1)^m \exp (-2 f^2 ) \sqrt{\frac{m!}{k!}} (2f )^{k-m}
 L^{k-m}_m ( 4f^2 ); \quad
k \geq m; \quad S_{km} = S_{mk}.
\end{eqnarray}

\noindent where  $L_m^k(x)$ are the  generalized  Laguerre polynomials  (the symmetrical matrix  $S_{mk}$ appears as the result of taking into account the exact integral of motion (\ref{7}) ).

Using  the same approximation, the expansion coefficients  of the wave function (\ref{8})  can be  derived as \cite{5}

\begin{eqnarray}
\label{11}
C^{np}_{l+} \approx B_{np}[\gamma \delta_{ln} +   \delta_{l,n+q}]; \
C^{np}_{l-} \approx p B_{np} [( \gamma  S_{nn} +   S_{n,n+q})\delta_{ln}
+( \gamma S_{n+q,n} +   S_{n+q,n+q}) \delta_{l,n+q}];\nonumber\\
 q = p(-1)^n; \ B^2_{np} = [(\gamma^2 + 1) + ( \gamma  S_{nn} +   S_{n,n+q})^2 + ( \gamma S_{n+q,n} +   S_{n+q,n+q})^2]^{-1}; \nonumber\\
\gamma = - \frac{p \Delta S_{n,n+q}}{2n - 2f^2
+ p\Delta S_{nn} - 2E_{np}} = - \frac{p \Delta S_{n,n+q}}
{ \frac{p\Delta}{2}(S_{nn} - S_{n+q, n+q}) - q\Omega  + q M},
\end{eqnarray}

\noindent where  $\delta_{kl}$ is the Kronecker delta, the factor $B_{np}$  is determined by the normalization condition  of the wave vector.

It was shown in \cite{13} that Eqs.(\ref{10}), (\ref{11}) can be used in the entire ranges of $f$ and $n$ values. However, according to \cite{8b},  the  main  applications of the considered problem are based on relatively weak interaction but intense driving field ($f \ll 1; \ n \gg 1, \ f\sqrt{n} \sim 1$). 
Usually the driving field is described by the coherent state \cite{Shirley,8b}. However, in the case of $n\gg 1$  fluctuations of the field amplitude $\Delta n \approx \sqrt{n} \ll n$ and the coherent state is appoximately equal to the Fock field state with the quantum number $n = \bar{n}$ \cite{Scully}.   As a result, all values in the Eqns. (\ref{10}) and (\ref{11}) depend only on the Rabi frequency (\ref{5a})  which is treated  as a constant value and  defined via  the average photon number $\bar{n}$   corresponding to the power $W$ of the driving field pulse and its duration   $\tau$  as

\begin{eqnarray}
\label{12}
\bar{n} = \frac{W \tau}{\hbar}; \ \Omega_R = 2 f \sqrt{\bar{n}}.
\end{eqnarray}

\section{Calculation of the quasi-energies and spontaneous emission for the sample model}
\label{sec.3}

The lifetime of the excited state   is defined by the interaction between  the atom  and  the VEF mode inside the cavity that leads to the transition of the atom to the  ground state accompanied by emitting a photon  $(\vecs{k}s)$. For applications it is important to consider the case when the frequency of the driving field $\Omega$ used for the lifetime control is essentially smaller than the spontaneous emission frequency $\omega_{\vecs{k}} \approx |E_0|$  \cite{2,3}. As a result. This means that in the result of the atomic transition to the ground state it does not interact with the driving field which  appears  in one of its stationary states with a photon number $m$. Thus, in the first-order approximation to  the atom-VEF interaction the wave function of the system final state is described by the following linear superposition:

\begin{eqnarray}
\label{13}
| \psi^{(1)}_{np} \rangle = | \Psi_{np} \rangle |0_b \rangle +
\sum_{m\vecs{k}s} B^{np}_{m\vecs{k}s}|\psi_{m\vecs{k}s}\rangle; \nonumber\\
|\psi_{m\vecs{k}s}\rangle = |\chi_0, m\rangle
b^+_{\vecs{k}s} |0_b \rangle;  \
|\chi_0, m\rangle = |\chi_0 \rangle \frac{1}{\sqrt{m!}}
(a^+)^m |0_a\rangle.
\end{eqnarray}

\noindent where $|0_b \rangle $ is the ground state of VEF and  the coefficients $B_{m\vecs{k}s}$   are  evaluated directly from (\ref{3}):

\begin{eqnarray}
\label{14}
B^{np}_{m\vecs{k}s} = - \frac{1}{m  + \omega_{\vecs{k}} - E_0 - E_{np} - i0}
\langle \chi_0, m | d_{\vecs{k}s} e^{-i\vecs{k}\vecs{r}}
|\Psi_{np}\rangle; \
d_{\vecs{k}s} =
- \frac{e}{m_e}
(\frac{2\pi}{V \omega_{\vecs{k}}})^{1/2}
(\vec{e}_{\vecs{k},s} \hat{\vec{p}}).
\end{eqnarray}

Calculating the second-order approximation for the quasi-energies with the state vector (\ref{13}), one can find in the  Weisskopf-Wigner  approximation \cite{10}:

\begin{eqnarray}
\label{15}
E_{np} = E^{(0)}_{np}
- \sum_{m\vecs{k}s} \frac{
\vert \langle \chi_0, m  | d_{\vecs{k}s} e^{-i\vecs{k}\vecs{r}}
| \Psi_{np} \rangle\vert^2
}{m  + \omega_{\vecs{k}} - E_0 - E_{np} - i0}.
\end{eqnarray}

In this approximation the imaginary part of the energy is assumed to be small compared  to   its  real part, the solution of this equation for one of the exited states of the "dressed" atom  can be obtained:

\begin{eqnarray}
\label{16}
E_{np} \simeq E^{(0)}_{np} + \delta E_{np} - i\frac{\Gamma_{np}}{2}; \nonumber\\
\Gamma_{np} =
2\pi \sum_{m\vecs{k}s}
\vert \langle \chi_0, m | d_{\vecs{k}s} e^{-i\vecs{k}\vecs{r}}
| \Psi_np \rangle\vert^2
\delta(m  + \omega_{\vecs{k}} - E_0 - E^{(0)}_{np}).
\end{eqnarray}

Applying  Eqn.(\ref{8}) to the wave function of the "dressed" atom, the matrix element of the operator of its interaction with   VEF can be expressed via the transition matrix elements of the "bare" atom:

\begin{eqnarray}
\label{17}
\langle \chi_0,m| d_{\vecs{k}s} e^{-i\vecs{k}\vecs{r}}
| \Psi_{np} \rangle
= \sum_{lq} C^{np}_{lq}
\langle \chi_0| d_{\vecs{k}s} e^{-i\vecs{k}\vecs{r}}
| \chi_q \rangle \langle m|l,f \rangle.
\end{eqnarray}

Using the dipole approximation for the matrix elements of the interaction with   VEF, one can use the pair of vectors proportional to the transition dipole momenta that define the "radiative" widths of the "bare" atom:

\begin{eqnarray}
\label{18}
\vec{d}_{\uparrow,\downarrow} =
\frac{e}{m_e}\langle\chi_0|\hat{\vec{p}}|\chi_{\uparrow,\downarrow}\rangle.
\end{eqnarray}

Then, after summing over the polarization and integrating over the wave vectors of the emitted photon in (\ref{17}), one can obtain:

\begin{eqnarray}
\label{19}
\Gamma_{np} =
\frac{2}{3} \sum_{m  < (E_{np} + E_0)}
(E^{(0)}_{np} + E_0 - m )\vert
\sum_{l} \langle l, f|m \rangle
[ C^{np}_{l+}(\vec{d}_{\uparrow} + \vec{d}_{\downarrow})
+ C^{np}_{l-}(\vec{d}_{\uparrow} - \vec{d}_{\downarrow})]\vert^2.
\end{eqnarray}

Equation  (\ref{19})  is derived  by usual approximations which are suitable for calculating radiative lifetimes in nonrelativistic cases. However one can see, that the radiative width represented by the sum of positively  defined  terms includes  both   the radiative widths of the excited states of a free atom and   the terms defined by the interfering matrix elements of both transitions.
An estimation for radiation width  valid  for an arbitrary amplitude of the field was obtained in \cite{7} . However, in the present paper we carry out an analytical computation directly  from  equation (\ref{19}) bounded by inequality (\ref{1}). In this case, the overlapping integral $\langle l, f|m \rangle$ can be obtained in the form \cite{13}

\begin{eqnarray}
\label{19a}
\langle l, f|m \rangle = \delta_{mn} + O ( \frac{\Omega^2_R }{\Omega^2}) = \delta_{mn} + O ( f^2 n ).
\end{eqnarray}

\noindent  so that Eqn.(\ref{19}) transforms as follows:

\begin{eqnarray}
\label{20}
\Gamma_{np} \approx
\frac{2}{3}
\sum_{m} (E^{(0)}_{np} + E_0 - m )
\vert[ C^{np}_{m +}(\vec{d}_{\uparrow} + \vec{d}_{\downarrow})
+ C^{np}_{m -}(\vec{d}_{\uparrow} - \vec{d}_{\downarrow})]\vert^2.
\end{eqnarray}

After some algebraic rearrangements we obtain:

\begin{eqnarray}
\label{21}
\Gamma_{np} \equiv
\frac{2}{3} [|\vec{d}_{\uparrow}|^2 D^{np}_{\uparrow} +
|\vec{d}_{\downarrow}|^2 D^{np}_{\downarrow} +
2Re(\vec{d}_{\uparrow}\vec{d}^*_{\downarrow}) D^{np}_{\pm}]; \nonumber\\
\Gamma_{\uparrow,\downarrow} =
\frac{4}{3} (E_0 \pm \Delta /2) \vert \vec{d}_{\uparrow,\downarrow}
\vert^2;
\nonumber\\
D^{np}_{\uparrow,\downarrow} = \sum_{m} (E^{(0)}_{np} + E_0 - m)\vert C^{np}_{m+} \pm
C^{np}_{m-}\vert^2; \qquad
D^{np}_{\pm} = \sum_{m}(E^{(0)}_{np} + E_0 - m) [|C^{np}_{m +}|^2 - |C^{np}_{m -}|^2].
\end{eqnarray}

The introduced  terms  $\Gamma_{\uparrow,\downarrow}$  coincide with excited state linewidths of a free atom. All other values  are  connected with the renormalization of the dipole  moments  of the "dressed" atom due to the interference in a superposition state  because of the atom-external field interaction.

\begin{figure}[tb]
\includegraphics[scale=0.6]{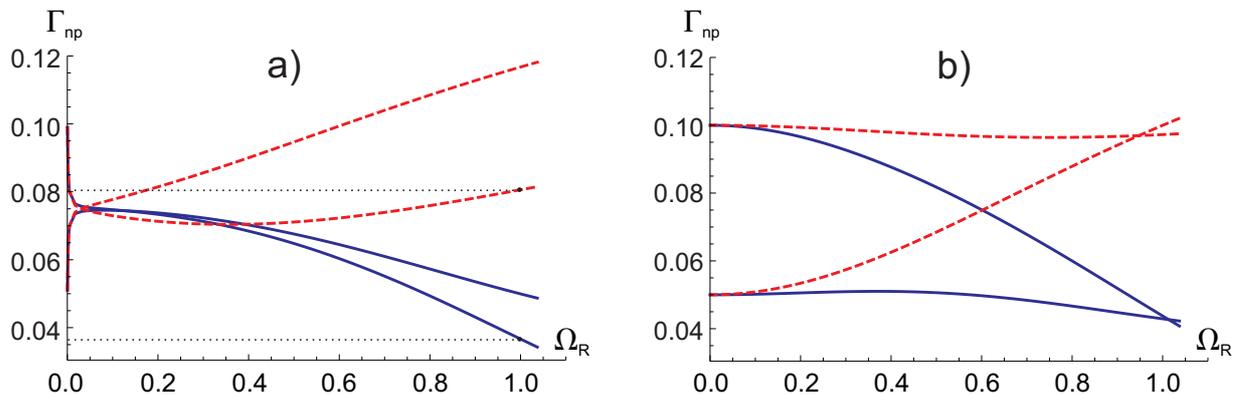}
\caption{The spectal linewidth as a function of Rabi-frequency for $n_0 = 225$, $\Gamma_\uparrow = 0.10$, $\Gamma_\downarrow = 0.05$ and $E_0 = 50$: (a) - the case of exact resonance ($\Delta = 1.0$), (b) - the case with detuning ($\Delta = 0.5$). Solid lines correspond to UAA and dashed lines to RWA. Dotted lines show the values of $\Gamma_{n-}$ at the Rabi frequency $\Omega_R = 1.0$ and resonance case for both UAA and RWA. }
\label{Fig.2}
\end{figure}

 In Fig.\ref{Fig.2}a and Fig.\ref{Fig.2}b  the linewidths calculated by means of RWA and UAA are compared  for different values of the system parameters. These values are chosen according to the conditions (\ref{2a}). These figures show noticeable difference in the behavior of the linewidths for these two cases depending on the Rabi frequency of the system.

The evolution of  such a  system is usually considered by means of the kinetic equation  for  the density matrix with relaxation term in the form of Lindblad super-operator\cite{14} added to it. This operator includes the fixed relaxation parameters  that is not valid in  the considered case because the relaxation dynamics of the systems depends on the interference of the transition matrix elements (\ref{21}). Therefore we use the Schr\"odinger picture for the state vector of the system  and expand it in terms of the quasi-stationary states of the system (\ref{13}) :

\begin{eqnarray}
\label{22}
|\Psi (t)\rangle = \sum_{np} G_{np} | \psi^{(1)}_{np} \rangle e^{- i E_{np}t} + \sum_{m\vecs{k}s} F_{m\vecs{k}s} |\psi_{m\vecs{k}s}\rangle e^{- i (\omega_{\vecs{k}} + m - E_{0})t}.
\end{eqnarray}

The expansion coefficients  $G_{np}, F_{m\vecs{k}s}$  should be  obtained  from the initial condition for the system state-vector $|\Psi (0)\rangle$.  In the analysis of the fluorescence spectrum, we  assume  that at initial  moment of  time the atom is in one of its excited quasi-stationary states while the ground state is empty. Then from the initial conditions the following connection between the coefficients can be deduced:

\begin{eqnarray}
\label{23}
F_{m\vecs{k}s} = - \sum_{np} G_{np} B^{np}_{m\vecs{k}s},
\end{eqnarray}
where $B^{np}_{m\vecs{k}s}$  is defined by (\ref{13}).

In the process of the spontaneous emission the atom transits to the ground state with emitting a photon. The probability amplitude of this process at time $t$ is defined by the term including the state vector   $|\psi_{m\vecs{k}s}\rangle$  in the expansion (\ref{22}):

\begin{eqnarray}
\label{24}
A_{m\vecs{k}s}(t) = \sum_{np} G_{np}B^{np}_{m\vecs{k}s}[ e^{- i E_{np}t} - e^{- i (\omega_{\vecs{k}} + m - E_{0})t}].
\end{eqnarray}

The spectrum of the spontaneous emission \cite{4} is defined by the modulus of the amplitude $A_{m\vecs{k}s}(\infty)$ squared, averaged over the initial states of the system  $G_{np}$  and summed over final states $m$ of the driving field.  Calculation of  the imaginary part of  $E_{np}$ should be taken into account and  in the limit $t \rightarrow \infty$ one can obtain

\begin{eqnarray}
\label{24}
S_{\vecs{k}}(\omega) = \sum_s \sum_m \sum_{np}\sum_{n'p}\rho^{n'p'}_{np}(B^{n'p'}_{m\vecs{k}s})^* B^{np}_{m\vecs{k}s}; \nonumber\\
\rho^{n'p'}_{np} = \overline{G_{np}G^{*}_{n'p'}},
\end{eqnarray}

\noindent where $\rho^{n'p'}_{np}$ is the density matrix of the atom  at the initial moment of time written  in terms of the excited "dressed" states.

In order to compare the  yields  of the spontaneous emission  spectra   described  in the framework of RWA and UAA, it is sufficient to consider the case of a stationary excitation which corresponds to the diagonal density matrix:

\begin{eqnarray}
\label{25}
\rho^{n'p'}_{np} = \rho_{np} \delta^{n'p'}_{np}.
\end{eqnarray}

In the "strong field"  case which corresponds to the normalized distribution $\rho_{np}$ having a sharp maximum near $n_0 \gg 1$  defined by the field amplitude, Eqn. (\ref{24}) can be written as:

\begin{eqnarray}
\label{26}
S_{\vecs{k}}(\omega) = \frac{1}{2}\sum_s \sum_{m =0}^{\infty} \sum_{p} | B^{n_0 p}_{m\vecs{k}s}|^2,
\end{eqnarray}

\noindent where it is assumed that the states with different   signs of the parity  $p = \pm 1$  are populated equally.
By using (\ref{14}) for $B^{np}_{m\vecs{k}s}$  in the dipole approximation and definition (\ref{16}) we obtain

\begin{eqnarray}
\label{27}
S_{\vecs{k}}(\omega) = \frac{1}{2}\sum_s \sum_{m =0}^{\infty}\sum_{p}   \frac{|\langle \chi_0, m | d_{\vecs{k}s}
|\Psi_{n_0 p}\rangle|^2}{[(m  + \omega_{\vecs{k}} - E_0 - E^{(0)}_{n_0 p})^2 + 1/4 \Gamma_{n_0 p}^2] },
\end{eqnarray}

\noindent where $\Gamma_{n_0 p}$  is defined by  (\ref{21}).

Performing the calculations  similar  to the ones in (\ref{19}), summing over the polarizations and calculating the density state of the emitted photon, one can derive the intensity of the spontaneous emission in solid angle $d \Omega$  across the direction of unit vector  $\vec n$ and within unitary spectral interval:

\begin{eqnarray}
\label{28}
\frac{\partial^2 S_{\vecs{k}}(\omega)}{\partial\Omega \partial \omega}  =
\frac{\omega}{(4 \pi)^2} \sum_{m =0}^{\infty}\sum_{p}
\frac{D^{(m)}_{\uparrow} A_{\uparrow} + D^{(m)}_{\downarrow} A_{\downarrow} + 2 D^{(m)}_{\pm} A_{\pm}}{[(m  + \omega  - E_0 - E^{(0)}_{n_0 p})^2 + 1/4 \Gamma_{n_0 p}^2] };\nonumber\\ D^{(m)}_{\uparrow,\downarrow} = \sum_{l,l'} \langle l, f|m \rangle \langle m|l', f \rangle
[ (C^{n_0p}_{l+}   \pm  C^{n_0p}_{l-} ) (C^{n_0p}_{l'+}   \pm  C^{n_0p}_{l'-} )]; \nonumber\\  D^{(m)}_{\pm} = \sum_{l,l'} \langle l, f|m \rangle \langle m|l', f \rangle [ C^{n_0p}_{l+} C^{n_0p}_{l'+}  -  C^{n_0p}_{l-}  C^{n_0p}_{l'-} ]; \nonumber\\
A_{\uparrow, \downarrow} = |\vec{d}_{\uparrow, \downarrow}|^2 - |(\vec n\vec{d}_{\uparrow, \downarrow})|^2; \ A_{\pm} =   \Re [(\vec{d}_{\uparrow} \vec d_{\downarrow}) - (\vec{d}_{\uparrow} \vec n) ( \vec d_{\downarrow}\vec n)].
\end{eqnarray}

When the coupling constant is sufficiently small, we can use the approximation (\ref{19a}) for the overlapping integral. Furthermore, to simplify the expression we consider the case of antiparallel dipole moment transitions, as it was depicted in Fig.1 and 2. Besides we assume that the emitted photon is observed in the direction perpendicular to  $\vec d$.  As a result, the final expression for the intensity distribution function can be derived:

\begin{eqnarray}
\label{29}
I_{\vecs{k}}(\omega) \equiv \frac{\partial^2 S_{\vecs{k}}(\omega)}{\partial\Omega \partial \omega}  =
\frac{\omega}{(4 \pi)^2} \sum_{m =0}^{\infty}\sum_{p}
\frac{1}{[(m  + \omega  - E_0 - E^{(0)}_{n_0 p})^2 + 1/4 \Gamma_{n_0 p}^2] } \times \nonumber\\
\times
[| d_{\uparrow}|^2 |C^{n_0p}_{m+} + C^{n_0p}_{m-}|^2 + |d_{\downarrow})|^2
|C^{n_0p}_{m+} - C^{n_0p}_{m-}|^2 - 2 \Re ( d_{\uparrow}d_{\downarrow})(|C^{n_0p}_{m+}|^2 - |C^{n_0p}_{m-}|^2)].
\end{eqnarray}

\begin{figure}[tb]
\includegraphics[scale=0.60]{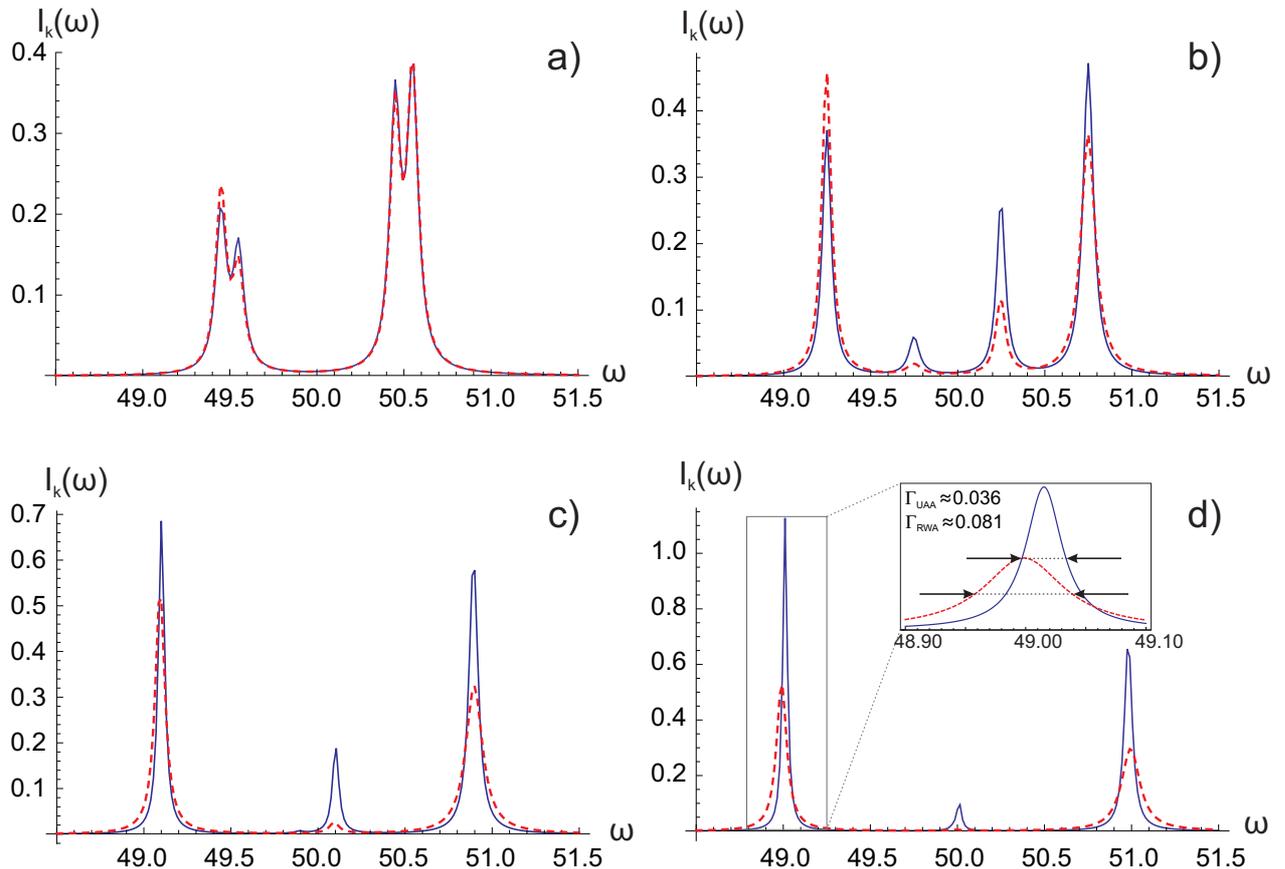}
\caption{The fluorescence spectra as a function of emitted photon frequency: (a) - $\Omega_R = 0.1$, (b) - $\Omega_R = 0.5$, (c) - $\Omega_R = 0.8$, (d) - $\Omega_R = 1.0$. Solid lines correspond to UAA and dashed lines to RWA. Inset in (d) shows the absolute values of the spectral linewidths at FWHM (see Fig.\ref{Fig.2}a for comparison).}
\label{Fig.4}
\end{figure}

Fig.\ref{Fig.4} shows the results of  estimation  of the fluorescence spectra $I_{\vecs{k}}(\omega)$  within the framework of RWA and UAA (the parameters used in the estimation are the same as in Fig.\ref{Fig.2}a for the case of exact resonance).

One can see from these figures that the positions of the peaks almost coincide for RWA and UAA because of the considered condition $(\Omega_R < \Omega)$. However,   widths and intensities of the peaks could differ essentially  for both cases because of $(\Gamma_{1,2} \ll \Omega_R)$. Asymmetry of the peak magnitudes are conditioned by the different values of the parameters $\Gamma_{\uparrow}$ and $\Gamma_{\downarrow}$ for the upper and lower levels of the atom in the considered model.

\section{Possible realization of the experiment}
\label{sec.4}

Let us discuss one of the possible realizations of a quantum system in order to observe the discussed phenomenon. As it was mentioned in \cite{2}, possibility to control the relaxation processes has general character and can be achieved by affecting atoms, nuclei and quantum nanostructures via the resonant field. At the same time, in order to control the system it is important to use electromagnetic field with the frequency much lower than the frequency of the spontaneous emission, in contrast to the case of resonant fluorescence \cite{Scully}. This condition arises from the analysis of a model system mentioned above. This analysis showed that essential influence of the resonant field on the spontaneous emission process takes place when the Rabi frequency of the system is comparable to the frequency of the resonant field.

If one considers  the atom-field resonant interaction for the natural atomic energy levels, the condition   $\Omega_R \ll \Omega$ takes place as a rule, even in the case of strong electromagnetic fields \cite{4}. However, the condition $\Omega_R  > \Omega$   can be fulfilled for nanostructures \cite{8,8a}.

\begin{figure}[tb]
\includegraphics[scale=0.5]{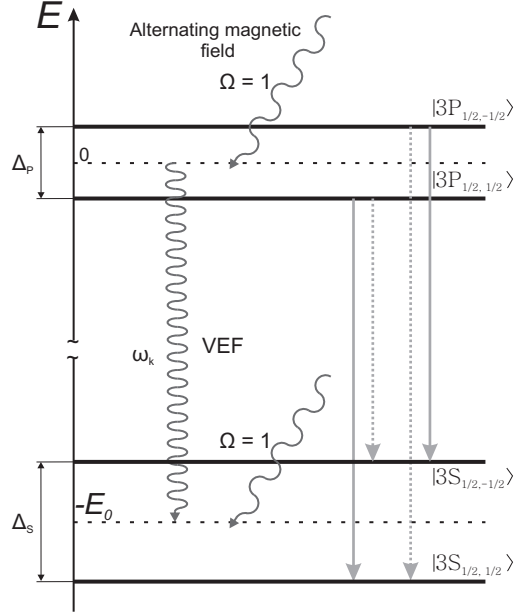}
\caption{The scheme of the transitions $3 P_{1/2} \rightarrow 3 S_{1/2}$ splitted in the field $B_z$ and exposed to the driving field  $B_x$. Solid lines correspond to emission of the photons with $\sigma$- polarization, dotted lines describe $\pi$-polarization. Parameters are defined in the text. }
\label{Fig.5}
\end{figure}

The conditions mentioned above can also be achieved by using the Zeeman splitting of the atomic energy levels in the stationary magnetic field $B_z$  and the driving alternating magnetic field with the amplitude $B_x$. Let us consider such a splitting of the atomic energy levels that correspond to $3 P_{1/2} \rightarrow 3 S_{1/2}$ transitions in sodium  (Fig.\ref{Fig.5}).

In contrast to the model system  (Fig.\ref{Fig.1}), it is important to take into account the interaction of the driving field with both excited and ground states of the sodium atom. These interactions are defined by the following Hamiltonians:

\begin{eqnarray}
\label{30a}
\hat H _{(P,S)} = - \frac{1}{2}g_{(P,S)} \mu_B \sigma_3 B_z + \frac{1}{2}g_{(P,S)} \mu_B \sigma_1 \hat B_x + \Omega a^+ a; \nonumber\\
\hat B_x =  B_{x0}(a + a^+).
\end{eqnarray}

\noindent where $\mu_B$ is the Bohr magneton; $g_P = 2/3; g_S = 2$ are the Lande factors for both states; $\Omega$ is the frequency of the quantum (driving) magnetic field and the amplitude $B_{x0}$ is chosen in such a way that for the case of the coherent state $|u \rangle$ of the field with the average number of quanta $\bar n = u^2$ its average number coincides with the amplitude $B_x$ of the classic alternating magnetic field:

\begin{eqnarray}
\label{30b}
\langle u|\hat B_x|u \rangle =  B_{x0} \langle u| a  |u \rangle =    B_{x0} \sqrt{\bar n} = B_x.
\end{eqnarray}

Thus, in order to construct the "dressed" states of the atom one should consider two separate equations similar to (\ref{6}):

\begin{eqnarray}
\label{30}
\{\frac{1}{2} \Delta_P \sigma_3 + \Omega a^+ a + f_P \sigma_1 (a + a^+)
- E^{(P)}_{np}\} | \Psi^{(P)}_{np} \rangle = 0; \nonumber\\
\{\frac{1}{2} \Delta_S \sigma_3 + \Omega a^+ a + f_S \sigma_1 (a + a^+)
- E^{(S)}_{np}\} | \Psi^{(S)}_{np} \rangle = 0,
\end{eqnarray}

\noindent where $\Delta_{(P,S)} = g_{(P,S)} \mu_B B_z$ and $f{(P,S)} \sim g_{(P,S)} \mu_B B_x$,  so that $\Delta_{S} = 3 \Delta_{P}; f_S = 3 f_P$. As it was described in Sec.\ref{sec.2}, in order to solve the equations (\ref{30}) within the framework of UAA it is convenient to use normalization of all parameters of the Hamiltonian on the frequency $\Omega$. In order to make the numerical simulation let us set the unity value corresponding to the resonance in the relatively weak magnetic field $B_0 = 10^2 \ Gs$. As a result, we obtain:

\begin{eqnarray}
\label{31}
\Omega = g_{P} \mu_B B_0 \rightarrow 1; \nonumber\\
\Delta_{S} = 3 \Delta_{P} = 3 \frac{B_z}{B_0}; \ f_P = 0.01; f_S = 0.03\ \Omega_R = 2 f_P \sqrt{n} = \frac{B_x}{B_0}.
\end{eqnarray}

Thus, the values of the Rabi frequency in the range $0.1 \Omega < \Omega_R <\Omega  $ correspond to the quantum excitations of the two-level system in the range $ 10^2 < n < 10^4$ and the amplitudes of the alternating magnetic field in the range $10 \ Gs < B_x < 100  \ Gs$.

In Sec.\ref{sec.3} the energy of the excited state ($3 P_{1/2}$ for the present case) is chosen as zero, so that in such energy units we get:

\begin{eqnarray}
\label{32}
E_0 = - \frac{E_{3 S_{1/2}}- E_{3 P_{1/2}}}{g_{P} \mu_B B_0} \approx 5 \times 10^6; \nonumber\\
E_{3 S_{1/2}}- E_{3 P_{1/2}} = \frac{2\pi c \hbar}{\lambda} = \frac{6.28 3.0 10^{10} 6.58 10^{-16}}{589.6} eV  \approx 2 eV;\nonumber\\ g_{P} \mu_B B_0 = \frac{2}{3} 5.79 10^{-9} 10^3 eV = 4 \times 10^{-7} eV.
\end{eqnarray}

In the considered case the interaction operator (\ref{5}) bounds two separate two-level systems (\ref{30}) together. In the selected direction of the axes of quantization the vectors of the atomic states correspond to the following configurations:

\begin{eqnarray}
\label{33}
|\Psi_{3P_{1/2}}, 1/2> = |R_{3,1}(r)>[-\sqrt{\frac{1}{3}}Y_{10} \chi_{1/2} + \sqrt{\frac{2}{3}} Y_{11} \chi_{-1/2}];\nonumber\\
|\Psi_{3P_{1/2}}, - 1/2> = |R_{3,1}(r)>[\sqrt{\frac{1}{3}}Y_{10} \chi_{-1/2} - \sqrt{\frac{2}{3}} Y_{1,-1} \chi_{1/2}]\nonumber;\\
|\Psi_{3S_{1/2}}, 1/2> = |R_{3,0}(r)> Y_{00} \chi_{1/2}; \nonumber\\
|\Psi_{3S_{1/2}}, - 1/2> = |R_{3,0}(r)> Y_{00} \chi_{- 1/2},
\end{eqnarray}

\noindent where $Y_{lm}$ are the spherical harmonics, $\chi_{\pm 1/2}$ are the spin functions of the valence electron.

The vectors of the dipole momenta of the selected transitions are defined by the matrix element of the operator $(- i \nabla)$. Their  amplitude $d_0$ depends on the radial wavefunctions as follows:

\begin{eqnarray}
\label{34}
d_0 = i \frac{e}{m_e} \int_0^{\infty}r^2  R_{3,1}(r) R'_{3,0}(r) dr ,
\end{eqnarray}
\noindent and its absolute value is not necessary for the present consideration.

However, directions of these vectors $\vec d_{s,s'} \ (s,s' = \pm 1/2)$ are essential

\begin{eqnarray}
\label{35}
    \vec d_{1/2,1/2} = \frac{1}{3} d_0 \vec e_z, \ \vec d_{-1/2,-1/2} = -\frac{1}{3} d_0 \vec e_z, \nonumber\\
    \vec d_{1/2,-1/2} = \frac{2}{3} d_0 (\vec e_x - i \vec e_y) = \frac{2}{3} d_0 \vec e_{-}, \
    \vec d_{-1/2,1/2} = \frac{2}{3} d_0 (\vec e_x + i \vec e_y) = \frac{2}{3} d_0 \vec e_{+}.
\end{eqnarray}

\noindent where $\vec e_j$ are the basis vectors of the chosen coordinate system.

Let us write the values of the natural width of the excited states of the atom in the absence of the alternating magnetic field as well (Fig.\ref{Fig.5}):

\begin{eqnarray}
\label{36}
\Gamma_1 \equiv \Gamma (3P_{1/2, 1/2}) = \Gamma_2 \equiv \Gamma (3P_{1/2, -1/2}) \equiv = \Gamma_0 = \frac{4 E_0}{3}|d_0|^2.
\end{eqnarray}

These values fulfill the condition (\ref{3}) by the order of magnitude ($Z$ is the charge of the nucleous, $a_0$ is the Bohr radius):

\begin{eqnarray}
\label{36a}
\Gamma \approx \frac{e^2 E_0^3 a_0^2}{Z^{2/3}} \ll \Omega_R,
\end{eqnarray}

\noindent that is essential in order to form the "dressed" states of the atom in the alternating magnetic field.

Let us consider the generalization of the expression (\ref{19}) for the width of the "dressed" excited states taking into account the retuning of the ground state in the alternating magnetic field. In this case the vector of the final state in (\ref{13}) is defined by the solution of the second equation in (\ref{30}):

\begin{eqnarray}
\label{37}
|\psi_{n' p'\vecs{k}s}\rangle = |\Psi_{n' p'}\rangle
b^+_{\vecs{k}s} |0_b \rangle.
\end{eqnarray}

As a result, expressions (\ref{16}), (\ref{17}) are transformed as follows:

\begin{eqnarray}
\label{38}
\Gamma_{np} =
2\pi \sum_{n' p'\vecs{k}s}
\vert \langle \Psi_{n' p'}| d_{\vecs{k}s} e^{-i\vecs{k}\vecs{r}}
| \Psi_{np} \rangle\vert^2
\delta(  \omega_{\vecs{k}} - E_0 + E^{(0)}_{n' p'} - E^{(0)}_{np}), \nonumber\\
\langle \Psi_{n' p'}| d_{\vecs{k}s} e^{-i\vecs{k}\vecs{r}}
| \Psi_{np} \rangle
=   \sum_{lq;l'q'} C^{n'p'(g)}_{l'q'} C^{np(e)}_{lq}
\langle \chi^{(g)}_{q'}| d_{\vecs{k}s} e^{-i\vecs{k}\vecs{r}}
| \chi^{(e)}_q \rangle \langle l',f_S|l,f_P \rangle.
\end{eqnarray}

\noindent where the index $(e,g)$ defines the coefficients and vectors of the excited and ground states of the system correspondingly.

Using dipole approximation in (\ref{38}) and applying condition (\ref{19a}) for the overlapping integral, one can deduce the following equation for the radiation width of the "dressed" atomic excited states:

\begin{eqnarray}
\label{39}
    \Gamma_{np} = \frac{4 E_0 |d_0|^2}{27} \sum_{n'p'} \left( \left|D^{n'p'-}_{np+} + D^{n'p'+}_{np-}\right|^2 +
         4 \left|D^{n'p'+}_{np+} - D^{n'p'-}_{np-}\right|^2 + 4 \left|D^{n'p'-}_{np+} - D^{n'p'+}_{np-}\right|^2 \right), \nonumber\\
    D^{n'p's'}_{nps} \equiv \sum_{l=0}^{\infty} C^{n'p'(g)}_{ls'} C^{np(e)}_{ls}.
\end{eqnarray}

\begin{figure}[tb]
\includegraphics[scale=0.6]{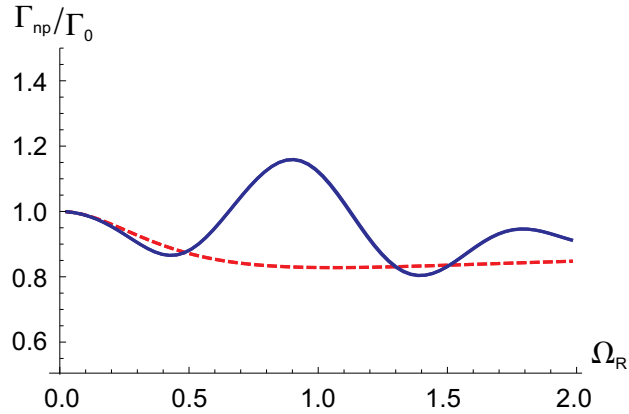}
\caption{The normalized spectral linewidth as a function of Rabi-frequency for $n_0 = 200$ for the case of exact resonance with $3 P_{1/2}$ state ($\Delta_P = 1.0$). Solid line  corresponds to UAA and dashed line  to RWA. }
\label{Fig.6}
\end{figure}

In Fig.\ref{Fig.6} the linewidth of the excited atomic "dressed" states normalized over the value (\ref{36}) is calculated for both RWA and UAA cases. One can conclude that there is a noticeable difference in the behavior of the linewidths for these two cases depending on the Rabi frequency of the system.

\section{CONCLUSIONS}

We have considered the possibility to control the natural spectral linewidth corresponding to the spontaneous transitions between excited and ground states of the atomic system exposed to the driving field. In contrast to the process of the resonant fluorescence, the frequency of the driving field was supposed to be several orders of magnitude less than the frequency of the spontaneous emission. It is shown that the noticeable change of the atomic excited state lifetime appears beyond the framework of RWA. The real physical system for which the effect could be observed is described.

\section{ACKNOWLEDGMENTS}

The authors are grateful to Professor S.Ya.Kilin for useful discussions. This work was supported by Alexander von Humboldt Foundation, Research Group Linkage Program and Belarusian State University under grant 200/1568 (05824).

{}

\end{document}